# Tsallis entropy: How unique?

Sumiyoshi Abe

*Institute of Physics, University of Tsukuba, Ibaraki 305-8571, Japan*

**Abstract** It is shown how, among a class of generalized entropies, the Tsallis entropy can uniquely be identified by the principles of thermodynamics, the concept of stability and the axiomatic foundation.

## 1. Introduction

The additivity requirement for the thermodynamic quantities puts constraint on symmetries of the phase/configuration space of the system under consideration: it is indivisibly connected with homogeneity of the system, which is a familiar *additional* assumption in ordinary thermodynamics . According to our experience, this requirement is not too stringent, and in fact it is fulfilled by ordinary thermodynamic systems.

Today, thermodynamically exotic complex systems/processes are attracting great attention. Examples include colossal magnetoresistance manganites, amorphous and glassy nanoclusters and high-energy collision processes. A common feature in these is that such systems stay in nonequilibrium stationary states for significantly long periods (compared to typical time scale of their microscopic dynamics) with preserving *scale* *invariant and hierarchical structures*. There, the phase/configuration spaces are generically inhomogeneous, and accordingly the naive additivity requirement may not be satisfied any more.

In this article, we discuss Tsallis' nonadditive entropy and show how it can uniquely be identified by the principles of thermodynamics, the concept of stability and the axiomatic foundation. In Sec. 2, a general composition law for thermodynamic entropy compatible with the zeroth law of thermodynamics is discussed. There, three generalized entropies, the Tsallis entropy, the normalized Tsallis entropy and the Rényi entropy, are shown to be consistent with the zeroth law. In Sec. 3, the concept of stability is examined for these generalized entropies, and it is shown that among the three only the Tsallis entropy is stable. In Sec. 4, the axioms and the uniqueness theorem for the Tsallis entropy are discussed. Sec. 5 is devoted to concluding remarks.

## 2. Composability

Suppose the total system be divided into two *independent* subsystems, $A$ and $B$. The joint probability of the total system is then factorized as $p_{ij}(A, B) = p_i(A)\, p_j(B)$, where $p_i(A)$ is the probability of finding the subsystem, $A$, in its $i$th state and so on. The concept of "composability" is stated as follows [1]. Entropy of the total system, $S(A, B) \equiv S[p(A, B)]$, is expressed in terms of entropies of the subsystems, $S(A)$ and $S(B)$. That is,

$$S(A, B) = f(S(A), S(B)), \tag{1}$$

where $f$ is a symmetric bivariate function

$$f(x, y) = f(y, x), \tag{2}$$

This function is referred to as the "composability function". The entropies frequently encountered in the literature satisfy the composability condition in eq. (1). The most celebrated one is the Boltzmann-Gibbs-Shannon entropy

$$S[p] = -\sum_{i=1}^{W} p_i \ln p_i, \tag{3}$$

where $W$ is the total number of microscopically accessible states at a given scale. This quantity is additive, i.e.,

$$S(A, B) = S(A) + S(B), \tag{4}$$

for independent $A$ and $B$. Another additive entropy is the Rényi entropy [2], which is given by

$$S_q^{(R)}[p] = \frac{1}{1-q} \ln \sum_{i=1}^{W} (p_i)^q \qquad (q > 0). \tag{5}$$

The Tsallis entropy [3]

$$S_q^{(T)}[p] = \frac{1}{1-q}\left[\sum_{i=1}^{W} (p_i)^q - 1\right] \qquad (q > 0) \qquad (6)$$

is also composable, since it satisfies

$$S_q^{(T)}(A, B) = S_q^{(T)}(A) + S_q^{(T)}(B) + (1-q) S_q^{(T)}(A)\, S_q^{(T)}(B), \qquad (7)$$

which is referred to as *pseudoadditivity*. The last example we mention here is the normalized Tsallis entropy [4,5]

$$S_q^{(NT)}[p] = \frac{S_q^{(NT)}[p]}{\sum_{i=1}^{W} (p_i)^q} = \frac{1}{1-q}\left[1 - \frac{1}{\sum_{i=1}^{W} (p_i)^q}\right] \qquad (q > 0), \qquad (8)$$

which also yields pseudoadditivity similar to eq. (7)

$$S_q^{(NT)}(A, B) = S_q^{(NT)}(A) + S_q^{(NT)}(B) + (q-1) S_q^{(NT)}(A)\, S_q^{(NT)}(B). \qquad (9)$$

We notice two important points. Firstly, the Rényi entropy, the Tsallis entropy and the normalized Tsallis entropy all converge to the Boltzmann-Gibbs-Shannon entropy in the limit $q \to 1$. Secondly, the Rényi entropy and the normalized Tsallis entropy are concave only for $q \in (0, 1)$, whereas the Tsallis entropy is always concave for any positive values of $q$.

Now, what is remarkable here is that the above entropies have a common feature, "sum plus product":

$$f(x, y) = x + y + \tau(q)\, xy, \qquad (10)$$

where $\tau(q) = 0$ for the Rényi entropy as well as the Boltzmann-Gibbs-Shannon entropy, $\tau(q) = 1 - q$ for the Tsallis entropy and $\tau(q) = q - 1$ for the normalized Tsallis entropy. In what follows, we shall see how this sum-plus-product structure is uniquely prescribed by the zeroth law of thermodynamics [6].

Let $X$ be an additive quantity of the system. It may be the internal energy, the volume or the number of particles, for example. Its total amount is given by

$$X(A, B) = X(A) + X(B). \qquad (11)$$

(It should be noticed that additivity of the internal energy is not obvious, in general. This makes the definition of temperature highly nontrivial for nonextensive systems like systems with long-range interactions. For the discussion about relaxing additivity of the internal energy, see Ref. [7].) Then, the stationary state of the total system may be defined by maximization of the total entropy with fixing $X(A, B)$. That is,

$$\delta S(A, B) = \delta f(S(A), S(B)) = 0, \qquad (12)$$

$$\delta X(A) = -\delta X(B). \tag{13}$$

From these, it follows that

$$\frac{\partial f(S(A), S(B))}{\partial S(A)} \frac{dS(A)}{dX(A)} = \frac{\partial f(S(A), S(B))}{\partial S(B)} \frac{dS(B)}{dX(B)}. \tag{14}$$

The zeroth law of thermodynamics requires this equation to be separated as

$$J(A) = J(B), \tag{15}$$

where $J$ is temperature, pressure or chemical potential if $X$ is the internal energy, the volume or the number of particles, respectively. Separation is realized if the derivatives of the composability function satisfy

$$\frac{\partial f(S(A), S(B))}{\partial S(A)} = k(S(A), S(B))\, g(S(A))\, h(S(B)), \tag{16}$$

$$\frac{\partial f(S(A), S(B))}{\partial S(B)} = k(S(A), S(B))\, h(S(A))\, g(S(B)), \tag{17}$$

where $k$, $g$ and $h$ are certain functions. It is not difficult to see that $k$ is a symmetric function, i.e., $k(x, y) = k(y, x)$, due to eq. (2).

We point out that, to establish eq. (15) with the correct physical dimensionality of $J$, $k$ may be dimensionless. It is very much likely that in most of ordinary physical systems $k$ is independent of the entropies of the subsystems. (A general case including an arbitrary symmetric function, $k$, can be found in Ref. [6].) Henceforth, $k$ is taken to be unity. Accordingly, eqs. (16) and (17) become

$$\frac{\partial f(S(A), S(B))}{\partial S(A)} = g(S(A))\, h(S(B)), \tag{18}$$

$$\frac{\partial f(S(A), S(B))}{\partial S(B)} = h(S(A))\, g(S(B)), \tag{19}$$

respectively.

Now, we impose the integrability condition

$$\frac{\partial^2 f(S(A), S(B))}{\partial S(A) \partial S(B)} = \frac{\partial^2 f(S(A), S(B))}{\partial S(B) \partial S(A)}. \tag{20}$$

Then, from eqs. (18) and (19), we obtain

$$\frac{1}{g(S(A))} \frac{d\, h(S(A))}{d\, S(A)} = \frac{1}{g(S(B))} \frac{d\, h(S(B))}{d\, S(B)} \equiv \tau, \tag{21}$$

where $\tau$ is a separation constant. Using this equation in eqs. (18) and (19), we have

$$\frac{\partial f_\tau(S(A), S(B))}{\partial S(A)} = \frac{1}{\tau}\frac{d h_\tau(S(A))}{d S(A)} h_\tau(S(B)), \tag{22}$$

$$\frac{\partial f_\tau(S(A), S(B))}{\partial S(B)} = \frac{1}{\tau} h_\tau(S(A)) \frac{d h_\tau(S(B))}{d S(B)}. \tag{23}$$

It turns out that the case $\tau = 0$ does not influence the final result (see below). Eqs. (22) and (23) are immediately integrated to yield

$$f_\tau(S(A), S(B)) = \frac{1}{\tau} h_\tau(S(A)) h_\tau(S(B)) + c, \tag{24}$$

where $c$ is an integration constant. This is the form of the composability function compatible with the zeroth law of thermodynamics.

Let us further determine the function, $h_\tau$. For this purpose, suppose both $A$ and $B$ be in the completely ordered states. In this case, $S(A) = S(B) = 0 = S(A, B)$. Then, eq. (24) gives

$$c = -\frac{1}{\tau} h_\tau^2(0). \tag{25}$$

Without losing generality, $h_\tau(0)$ can be set equal to unity

$$h_\tau(0) = 1. \tag{26}$$

Next, suppose only $B$ be in the completely ordered state. Then, $S(A, B) = S(A)$, and accordingly eq. (24) combined with eqs. (25) and (26) leads to

$$h_\tau(S) = \tau S + 1. \tag{27}$$

Substituting this into eq. (24), we find

$$S(A, B) = f(S(A), S(B)) = S(A) + S(B) + \tau S(A) S(B). \tag{28}$$

This precisely reproduces pseudoadditivity in eqs. (7) and (9).

Thus, we have shown that validity of the zeroth law of thermodynamics puts a stringent constraint on the composition law for entropy and the resulting law is pseudoadditivity, i.e., the sum-plus-product structure.

## 3. Stability

In the previous section, we have seen how the Rényi entropy, the Tsallis entropy and the normalized Tsallis entropy (and the Boltzmann-Gibbs-Shannon entropy as the limiting case) are compatible with composability and existence of the stationary state, i.e., the generalized-maximum-entropy state defined consistently with the zeroth law of thermodynamics. In this section, we examine if these generalized entropies can be physically observable quantities. For this purpose, we discuss the concept of stability introduced in Ref. [8].

Consider a quantity $C = C[p]$ with its maximum value, $C_{\max}$. What is experimentally observed is usually not $C$ but the distribution $\{p_i\}_{i=1,2,\cdots,W}$. Repeating measurement of a certain physical random variable (e.g., energy), one obtains a set of distributions, which do not exactly coincide with each other. If $C$ is a physically meaningful quantity, it should not drastically change under small change of the distribution. Otherwise, its values cannot be experimentally reproducible with reliability. Mathematically, this may be expressed as follows [8]:

$$(\forall \varepsilon > 0)\ (\exists \delta > 0)\ \left( \| p - p' \|_1 \le \delta \Rightarrow \left| \frac{C[p] - C[p']}{C_{\max}} \right| < \varepsilon \right) \tag{29}$$

for arbitrarily large values of $W$, where $\|A\|_1$ stands for the $l^1$ norm of $A$: $\|A\|_1 = \sum_{i=1}^{W} |A_i|$.

Let us examine stability of the generalized entropies in eqs. (5), (6) and (8). Note that all those entropies take their maximum values for the equiprobability $p_i = 1/W$ $(i = 1, 2, \cdots, W)$: $S_{q,\max}^{(R)} = \ln W$, $S_{q,\max}^{(T)} = \ln_q W$, $S_{q,\max}^{(NT)} = -\ln_q W^{-1}$, where $\ln_q x$ stands for the $q$-logarithmic function defined by $\ln_q x = (x^{1-q} - 1)/(1-q)$ $(x > 0)$ that tends to the ordinary logarithmic function, $\ln x$, in the limit $q \to 1$.

The following specific deformations of the distribution should be considered here:

(i) $0 < q < 1$;

$$p_i = \delta_{i1}, \qquad p'_i = \left(1 - \frac{\delta}{2}\frac{W}{W-1}\right)p_i + \frac{\delta}{2}\frac{1}{W-1}. \tag{30}$$

(ii) $q > 1$;

$$p_i = \frac{1}{W-1}(1-\delta_{i1}), \qquad p'_i = \left(1 - \frac{\delta}{2}\right)p_i + \frac{\delta}{2}\delta_{i1}. \tag{31}$$

These deformations are of physical importance in view of relaxation processes. In both cases, the $l^1$ norms are calculated to be $\| p - p' \|_1 = \delta$. It is immediate to evaluate the changes of the Rényi entropy, the Tsallis entropy and the normalized Tsallis entropy under these deformations. The results in the thermodynamic limit, $W \to \infty$, are given as follows:

(i) $0 < q < 1$;

$$\left| \frac{S_q^{(R)}[p] - S_q^{(R)}[p']}{S_{q,\max}^{(R)}} \right| = \left| \frac{\frac{1}{1-q} \ln\left[\left(1-\frac{\delta}{2}\right)^q + \left(\frac{\delta}{2}\right)^q (W-1)^{1-q}\right]}{\ln W} \right|$$

$$\to 1 \qquad (W \to \infty), \tag{32}$$

$$\left| \frac{S_q^{(T)}[p] - S_q^{(T)}[p']}{S_{q,\max}^{(T)}} \right| = \left| \frac{\left(1-\frac{\delta}{2}\right)^q + \left(\frac{\delta}{2}\right)^q (W-1)^{1-q} - 1}{W^{1-q} - 1} \right|$$

$$\to \left(\frac{\delta}{2}\right)^q \qquad (W \to \infty), \tag{33}$$

$$\left|\frac{S_q^{(NT)}[p]-S_q^{(NT)}[p']}{S_{q,\max}^{(NT)}}\right|=\left|\frac{1-\dfrac{1}{\left(1-\dfrac{\delta}{2}\right)^q+\left(\dfrac{\delta}{2}\right)^q(W-1)^{1-q}}}{1-W^{q-1}}\right|$$

$$\to 1 \qquad (W\to\infty). \tag{34}$$

(ii) $q>1$;

$$\left|\frac{S_q^{(R)}[p]-S_q^{(R)}[p']}{S_{q,\max}^{(R)}}\right|=\left|\frac{\ln(W-1)-\dfrac{1}{1-q}\ln\left[\left(\dfrac{\delta}{2}\right)^q+\left(1-\dfrac{\delta}{2}\right)^q(W-1)^{1-q}\right]}{\ln W}\right|$$

$$\to 1 \qquad (W\to\infty), \tag{35}$$

$$\left|\frac{S_q^{(T)}[p]-S_q^{(T)}[p']}{S_{q,\max}^{(T)}}\right|=\left|\frac{(W-1)^{1-q}-\left(\dfrac{\delta}{2}\right)^q-\left(1-\dfrac{\delta}{2}\right)^q(W-1)^{1-q}}{W^{1-q}-1}\right|$$

$$\to \left(\frac{\delta}{2}\right)^q \qquad (W\to\infty), \tag{36}$$

$$\left|\frac{S_q^{(NT)}[p]-S_q^{(NT)}[p']}{S_{q,\max}^{(NT)}}\right|=\left|\frac{-(W-1)^{q-1}+\dfrac{1}{\left(\dfrac{\delta}{2}\right)^q+\left(1-\dfrac{\delta}{2}\right)^q(W-1)^{1-q}}}{1-W^{q-1}}\right|$$

$$\to 1. \qquad (W\to\infty). \tag{37}$$

It should be noticed that although the three generalized entropies are related to each other, for example, as $S_q^{(T)} = (1-q)^{-1}\{\exp[(1-q)S_q^{(R)}]-1\} = S_q^{(NT)}/[1+(q-1)S_q^{(NT)}]$, their stability properties are thus different. The Rényi entropy and the normalized Tsallis entropy do not satisfy the stability condition and therefore cannot be used for generalizing statistical mechanics. These entropies with $0 < q < 1$ overestimate a large number of occupied states even if their overall probability is irrelevantly small and these with $q > 1$ simply overestimate a high peak of probability. *Accordingly, the Rényi entropy and the normalized Tsallis entropy cannot describe relaxation processes in a physically meaningful manner.* (We also mention that although instability of the Rényi entropy has already been shown in Ref. [8] we have included it here for the sake of completeness of our discussion.) Among the three, only the Tsallis entropy satisfies the stability condition in eq. (29) under the above-mentioned specific (but important) deformations if $\delta$ is taken to be $\delta < 2\varepsilon^{1/q}$.

In what follows, we shall present a proof that the Tsallis entropy is in fact stable under an arbitrary deformation of the probability distribution.

## 3. Stability of Tsallis entropy

Let us consider

$$A_q[p;t) = \sum_{i=1}^{W} \left( p_i - \frac{1}{e_q(t)} \right)_+ , \tag{38}$$

where $t$ is a positive parameter and $(x)_+ \equiv \max\{x, 0\}$. It is possible to show that the following basic inequalities hold:

$$\left| A_q[p;t) - A_q[p';t) \right| \le \| p - p' \|_1 , \tag{39}$$

$$\left| A_q[p;t) - A_q[p';t) \right| < \frac{W}{e_q(t)} \qquad (\forall t \ge \ln_q W). \tag{40}$$

A point is that the Tsallis entropy can be expressed in terms of $A_q[p,t)$ through the following integral representation [9]:

$$S_q^{(T)}[p] = -1 + q \int_0^{t_{\max}} dt \left(1 - A_q[p;t)\right), \tag{41}$$

where $t_{\max} = \infty$ if $0 < q < 1$ and $t_{\max} = 1/(q-1)$ if $q > 1$.

Using eq. (41), we find

$$\left| S_q^{(T)}[p] - S_q^{(T)}[p'] \right| \le q \int_0^{a+\ln_q W} dt \left| A_q[p;t) - A_q[p';t) \right|$$

$$+ q \int_{a+\ln_q W}^{t_{\max}} dt \left| A_q[p;t) - A_q[p';t) \right| , \tag{42}$$

where $a$ is a positive constant satisfying $-\ln_q W < 0 < a < t_{\max}$. From eq. (39), the first term on the right-hand side of eq. (42) is shown to satisfy

$$q \int_0^{a+\ln_q W} dt \left| A_q[p;t) - A_q[p';t) \right| \le q \, \| p - p' \|_1 \, (a + \ln_q W), \tag{43}$$

whereas from eq. (40) the second term is found to be

$$q \int_{a+\ln_q W}^{t_{\max}} dt \left| A_q[p;t) - A_q[p';t) \right| < \frac{W}{\left[ e_q\left(a + \ln_q W\right) \right]^q}. \tag{44}$$

These enable us to rewrite eq. (42) as follows:

$$\left| S_q^{(T)}[p] - S_q^{(T)}[p'] \right| < q \, \| p - p' \|_1 \, (a + \ln_q W) + \frac{W}{\left[ e_q\left(a + \ln_q W\right) \right]^q}. \tag{45}$$

Evaluating the right-hand side of this inequality, one finds that it takes its minimum value when $a = -\ln_q W + \ln_q(W / \| p - p' \|_1)$. Finally, replacing the right-hand side by its minimum value and recalling $S_{q,\max}^{(T)} = \ln_q W$, eq. (45) turns out to yield

$$\left| \frac{S_q^{(T)}[p] - S_q^{(T)}[p']}{S_{q,\max}^{(T)}} \right| < (\| p - p' \|_1)^q \left( 1 + \frac{1 - q \ln_q \| p - p' \|_1}{\ln_q W} \right)$$

$$\rightarrow \begin{cases} (\| p - p' \|_1)^q & (0 < q < 1) \\ q \| p - p' \|_1 & (q > 1) \end{cases} \qquad (W \rightarrow \infty). \qquad (46)$$

Therefore, taking $\delta$ such that $\| p - p' \|_1 \leq \delta < \varepsilon^{1/q}$ for $0 < q < 1$ and $\| p - p' \|_1 \leq \delta < \varepsilon / q$ for $q > 1$, the stability condition in eq. (29) is found to be fulfilled by the Tsallis entropy.

The above discussion is valid for $\forall q > 0$, and accordingly stability of the Boltzmann-Gibbs-Shannon entropy [8] (recovered in the limit $q \rightarrow 1$) is also reestablished as a by-product.

## 4. Axioms and uniqueness theorem

It is well known that the Boltzmann-Gibbs-Shannon entropy is uniquely identified by a set of axioms termed the Shannon-Khinchin axioms [10,11]. In this section, we briefly mention that the Tsallis entropy has a similar axiomatic foundation. More detailed discussions can be found in Refs. [12,13].

The set of axioms presented in Ref. [13] is the following: [I] $S_q(p_1, p_2, \cdots, p_W)$ is continuous with respect to all its arguments and takes its maximum for the equiprobability distribution $p_i = 1/W$ $(i = 1, 2, \cdots, W)$, [II] $S_q(A, B) = S_q(A) + S_q(B \mid A)$ $+(1-q) S_q(A) S_q(B \mid A)$ for a bipartite system, $(A, B)$, and [III] $S_q(p_1, p_2, \cdots, p_W, p_{W+1} = 0) = S_q(p_1, p_2, \cdots, p_W)$. It can be shown [13] that the quantity $S_q$ satisfying [I]-[III] is, up to a multiplicative constant, uniquely given by

$$S_q(p_1, p_2, \cdots, p_W) \equiv S_q^{(T)}[p] = \frac{1}{1-q}\left[\sum_{i=1}^{W} (p_i)^q - 1\right], \tag{47}$$

which is precisely the Tsallis in eq. (6). By comparison of this set of axioms with that of Khinchin [11], the one and only difference is seen to be in [II]. Khinchin's second axiom is recovered in the limit $q \to 1$. $S_q(B \mid A)$ in [II] is the nonadditive conditional entropy defined by

$$S_q(B \mid A) = < S_q(B \mid A_i) >_q^{(A)}, \tag{48}$$

provided that $S_q(B \mid A_i)$ is calculated from of the conditional probability of $B$ with $A$ found in its $i$th state, $p_{ij}(B \mid A) = p_{ij}(A, B) / p_i(A)$ with the marginal probability $p_i(A) = \sum_j p_{ij}(A, B)$. The symbol $< Q >_q^{(A)}$ stands for the normalized $q$-expectation value [14] defined by

$$< Q >_q^{(A)} = \sum_i Q_i P_i(A) \ , \tag{49}$$

where

$$P_i(A) = \frac{[p_i(A)]^q}{\sum_i [p_i(A)]^q} \tag{50}$$

is the escort distribution [15] associated with $p_i(A)$.

To avoid confusion, here we emphasize the following point. At the level of eq. (48), $S_q(B \mid A_i)$ is not specified yet. As a functional, this quantity should satisfy the axioms [I]-[III] for the conditional probability. After having reached at the result in eq. (47), overall consistency has to be ascertained. This can be done, for example, as follows. Calculate eq. (47) for the joint probability, $p_{ij}(A, B) = p_i(A)\, p_{ij}(B \mid A)$, to obtain

$$
\begin{aligned}
S_q^{(T)}(A, B) &= \frac{1}{1-q}\left\{\sum_{i,j} [p_i(A)\, p_{ij}(B \mid A)]^q - 1\right\} \\
&= \sum_i [p_i(A)]^q\, S_q^{(T)}(B \mid A_i) + S_q^{(T)}(A) \\
&= S_q^{(T)}(B \mid A) \sum_i [p_i(A)]^q + S_q^{(T)}(A) \\
&= S_q^{(T)}(A) + S_q^{(T)}(B \mid A) + (1-q)\, S_q^{(T)}(A) S_q^{(T)}(B \mid A),
\end{aligned}
\tag{51}
$$

which is consistent with [II]. Also, notice that there exists the following correspondence relation between the Bayes law and the generalized composition law in [II]:

$$p_{ij}(A, B) = p_i(A)\, p_{ij}(B \mid A) = p_j(B)\, p_{ij}(A \mid B)$$

$\leftrightarrow$

$$
\begin{aligned}
S_q(A, B) &= S_q(A) + S_q(B \mid A) + (1-q) S_q(A) S_q(B \mid A) \\
&= S_q(B) + S_q(A \mid B) + (1-q) S_q(B) S_q(A \mid B),
\end{aligned}
\tag{52}
$$

which naturally generalizes that of the Boltzmann-Gibbs-Shannon entropy realized in the limit $q \to 1$: $p_{ij}(A, B) = p_i(A)\, p_{ij}(B \mid A) = p_j(B)\, p_{ij}(A \mid B) \leftrightarrow S(A, B) = S(A) + S(B \mid A)$ $= S(B) + S(A \mid B)$. In [II], we clearly see the sum-plus-product structure discussed in Sec. 2. It is of importance to notice that this generalized composition law is more general than pseudoadditivity since it holds even when there is correlation between $A$ and $B$ [16].

Finally, we wish to point out that the set of axioms for the Tsallis entropy presented in this section may provide a basis for nonadditive information theory (and its variants [17]). Nowadays it has further been generalized to quantum theory. In particular, resulting nonadditive quantum information theory has recently been discussed in the context of the problems of quantum entanglement, extensively [18-25].

## 5. Concluding remarks

We have discussed the distinguished properties of the Tsallis entropy from three viewpoints: the principles of thermodynamics, the concept of stability and the axiomatic foundation. In particular, it is emphasized that the Tsallis entropy can provide a basis for generalizing statistical mechanics, whereas the Rényi entropy and the normalized Tsallis entropy are ruled out for such a purpose. Generalized statistical mechanics based on the Tsallis entropy, termed nonextensive statistical mechanics, is now under vital investigation. It has successfully been applied to a variety of complex systems. Its statistical and dynamical foundations have also been well developed. The references relevant to these developments can be found at URL [26].

## Acknowledgments

The author would like to thank Professor M. Sugiyama for giving him this opportunity to summarize the basic properties of the generalized entropies. He also thanks A. K. Rajagopal and C. Tsallis for discussions.